\documentclass[
    ,final            % use final for the camera ready runs
  ]
  {aipproc}

\layoutstyle{6x9}

\newcommand\simlt{\lower.5ex\hbox{$\; \buildrel < \over \sim \;$}}
\newcommand\simgt{\lower.5ex\hbox{$\; \buildrel > \over \sim \;$}}

\begin{document}

\title{On the origin of very high energy $\gamma$-rays from radio galaxies}

\classification{98.54.Gr, 95.30.Qd, 95.85.Pw, 98.62.En, 98.62.Nx}
\keywords{Radio galaxies, gamma-rays, jets, black holes, physical processes}
\author{Frank M. Rieger}{
  address={Max-Planck-Institut f\"ur Kernphysik, P.O. Box 103980, 69029 Heidelberg, Germany}
}

\begin{abstract}
Radio galaxies have emerged as a new gamma-ray emitting source class on the 
extragalactic sky. With their jets misaligned, i.e. not directly pointing towards us, 
they offer a unique tool to probe some of the fundamental (and otherwise hidden) 
non-thermal processes in AGN. This contribution briefly summarizes the observed
characteristics of the four radio galaxies detected so far at very high energies (VHE). 
Given its prominence, particular attention is given to the origin of the {\it variable} 
VHE emission in M87. We discuss some of the theoretical progress achieved for 
this source within recent years highlighting, amongst others, the relevance of 
magnetospheric particle acceleration and emission models.
\end{abstract}

\maketitle

%%%%%%%%%%%%%%%%%%%%%%%%%%%%%%%%%%%%%%%%%%%%
%% MAINMATTER
%%%%%%%%%%%%%%%%%%%%%%%%%%%%%%%%%%%%%%%%%%%%

\section{Introduction}
The extragalactic sky as observed at very high energies (VHE; $>100$ GeV) is 
highly dominated by Active Galactic Nuclei (AGNs) of the blazar-type, i.e., AGNs 
with relativistic jets pointing almost directly towards us. Flux enhancement 
of the intrinsic non-thermal jet emission by relativistic Doppler boosting effects 
then naturally favours their detection at VHE. On the other hand, misaligned (i.e., 
non-blazar) AGNs, characterized by jets substantially inclined with respect to
the observer, could allow us to probe some of the fundamental non-thermal 
VHE processes in AGN that are otherwise swamped by boosted jet emission.
Nearby radio galaxies (RGs) are especially attractive as their proximity often 
makes it possible to resolve their radio jets down to sub-parsec scales, and to 
investigate the behavior of the emission at different wavelengths.

\section{Radio Galaxies at gamma-ray energies}
At high gamma-ray energies (HE; $>100$ MeV) Fermi-LAT has reported the 
detection of about ten misaligned RGs, mostly of the Fanaroff-Riley-type~I 
(FR~I) \cite{Abdo10a,Ackermann11,Nolan12}, see also Fig. 1. 
At very high energies (VHE), only four RGs have been clearly identified so far: 
Cen~A ($d\simeq 3.8$ Mpc), M87 ($\simeq 16.7$ Mpc) and the Perseus Cluster 
($d\sim77$ Mpc, $z\sim 0.018$) sources NGC~1275 and IC~310. A detection 
of the RG 3C66B at VHE was reported by MAGIC (2007 observations \cite{Aliu09}), 
but the VHE emission seems not sufficiently disentangled from the nearby 
(separation $\theta \sim 0.12^{\circ}$) IBL blazar 3C66A to consider it here as a 
confirmed VHE emitter.\\
%%%%%%%%%%%%    FIGURE 1 %%%%%%%%%%%%%%%%%%
\begin{figure}
 \includegraphics[height=0.5\textheight]{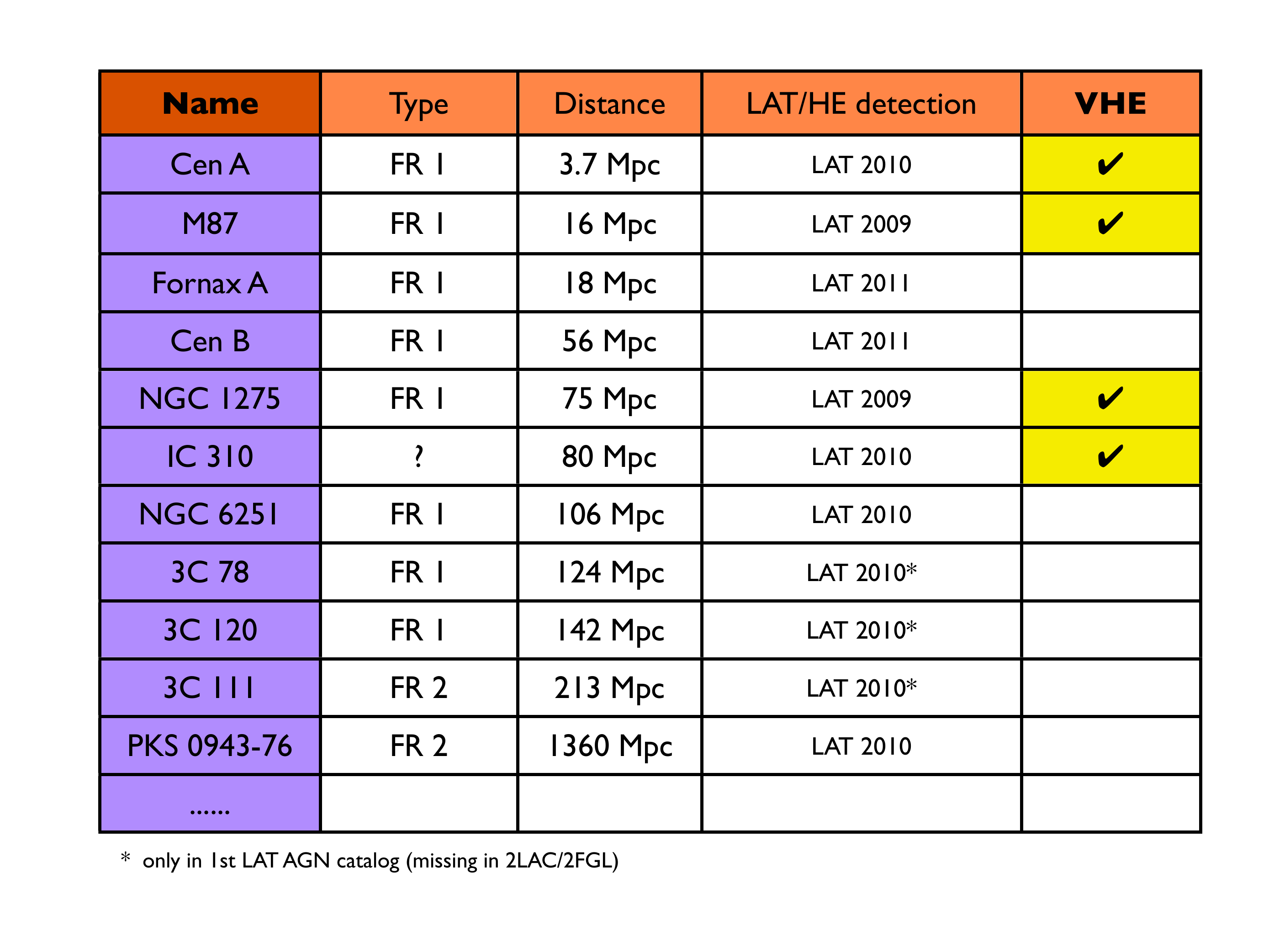}
  \caption{ Radio Galaxies (RGs) detected at gamma-ray energies. 
  Out of $\geq$ 886 AGN ("clean" sample) detected by Fermi at high
  energies (HE) less than a few percent are RGs. Only four of these 
  RGs are currently confirmed to be VHE emitters.}
\end{figure}
%%%%%%%%%%%%%%%%%%%%%%%%%%%%%%%%%%%%

\noindent The nearest AGN {\it Cen A}, hosting a black hole of mass 
$M_{\rm BH} \simeq (0.5-1) \times 10^8\,M_{\odot}$, was the second RG 
to be detected at VHE energies in a deep ($>120$h) exposure by H.E.S.S.
\cite{Aharonian09}. Cen~A is a weak VHE source with an integral flux 
$I(>250$ GeV) above 250 GeV of about $0.8\%$ of the Crab Nebula 
(equivalent to an apparent isotropic VHE luminosity of $L(>250$ GeV) 
$\simeq 2 \times 10^{39}$ erg/s). The VHE spectrum extends up to 
$\sim5$ TeV and is compatible with a single power law of photon index 
$2.7\pm0.5$. No significant VHE variability has been found so far, so 
that an extended origin of the VHE emission \cite{Hardcastle} cannot 
be simply excluded yet.  At high energies, Fermi-LAT has also detected 
HE $\gamma$-rays up to 10 GeV from the core of Cen~A \cite{Abdo10b}. 
The HE light curve (15d bins) of the core is consistent with no variability 
and the HE $\gamma$-ray spectrum can be described by a photon index 
comparable to the one in the VHE. A simple power law extrapolation of 
the HE spectrum, however, cannot account for the observed TeV flux. This 
could be indicative of an additional emission contribution to the VHE domain 
beyond the common synchrotron-Compton emission \cite{Chiaberge01}, 
emerging at the highest energies \cite{Rieger12a}. At GeV energies, Fermi-LAT 
has also detected $\gamma$-ray emission from the giant radio lobes of Cen A.
A recent analysis of a larger HE data set revealed significant evidence for 
a spatial extension of the $\gamma$-ray lobes beyond the radio (WMAP)  
image \cite{Zhang12}, suggesting the need for a more complex ($\geq$ 
two-zone) lobe emission modeling than done so far. Note that the angular 
resolution of H.E.S.S. ($\sim0.1^{\circ}$) clearly allows to exclude the lobe 
regions as source of the detected TeV emission.\\
The Virgo Cluster galaxy {\it M87} was the first radio galaxy detected at 
TeV energies \cite{Aharonian03}. Classified as an FR I, M87 is known to 
host one of most massive black holes with $M_{\rm BH} \simeq(2-6)\times
10^9\,M_{\odot}$ and to exhibit a relativistic jet misaligned by $\theta
\simeq(15-25)^{\circ}$. These latter values are consistent with modest 
Doppler boosting $D=1/[\Gamma_j (1-\beta\cos\theta)] \simlt 3$ (for review, 
see e.g. \cite{Rieger12b}). M87 has shown particularly interesting VHE 
features within recent years, including rapid day-scale variability (observed 
flux doubling time scales $\Delta t_{\rm obs} \sim 1$) above 350 GeV 
during active source states (with flux levels sometimes exceeding $10\%$ 
of the Crab Nebula), and a hard spectrum (compatible with a power law of 
photons index $\Gamma \simeq 2.2$) extending beyond $10$ TeV 
\cite{Aharonian06,Albert10,Aliu12,Abramowski12,Beilicke12}.
Both the hard VHE spectrum and the observed rapid VHE variability are 
remarkable features for a misaligned AGN. M87 is the only RG where 
clear evidence for day-scale TeV variability (see Fig. 2) has been found, 
although there appear to be some hints for day-scale variability in IC~310 
(and for month-type in NGC~1275) as well (see below). The observed 
variations in M87 seen at VHE are the fastest compared to any other 
waveband so far, and imply a compact $\gamma$-ray emitting region 
($R < c\Delta t_{\rm obs} D$), comparable to the Schwarzschild radius 
$r_s=(0.6-1.8) \times10^{15}$ cm of its black hole. This could point to the 
black hole vicinity as the most likely origin of the observed, variable VHE 
radiation, a consideration supported by high-resolution radio (VLBA) 
observations during the 2008 (February) VHE high state of M87, showing 
that the VHE outburst was followed by an ejection of a new radio component 
from very close to the black hole \cite{Acciari09}.
%%%%%%%%%%%%   FIGURE 2  %%%%%%%%%%%%%%%%%
\begin{figure}
  \includegraphics[height=0.4\textheight]{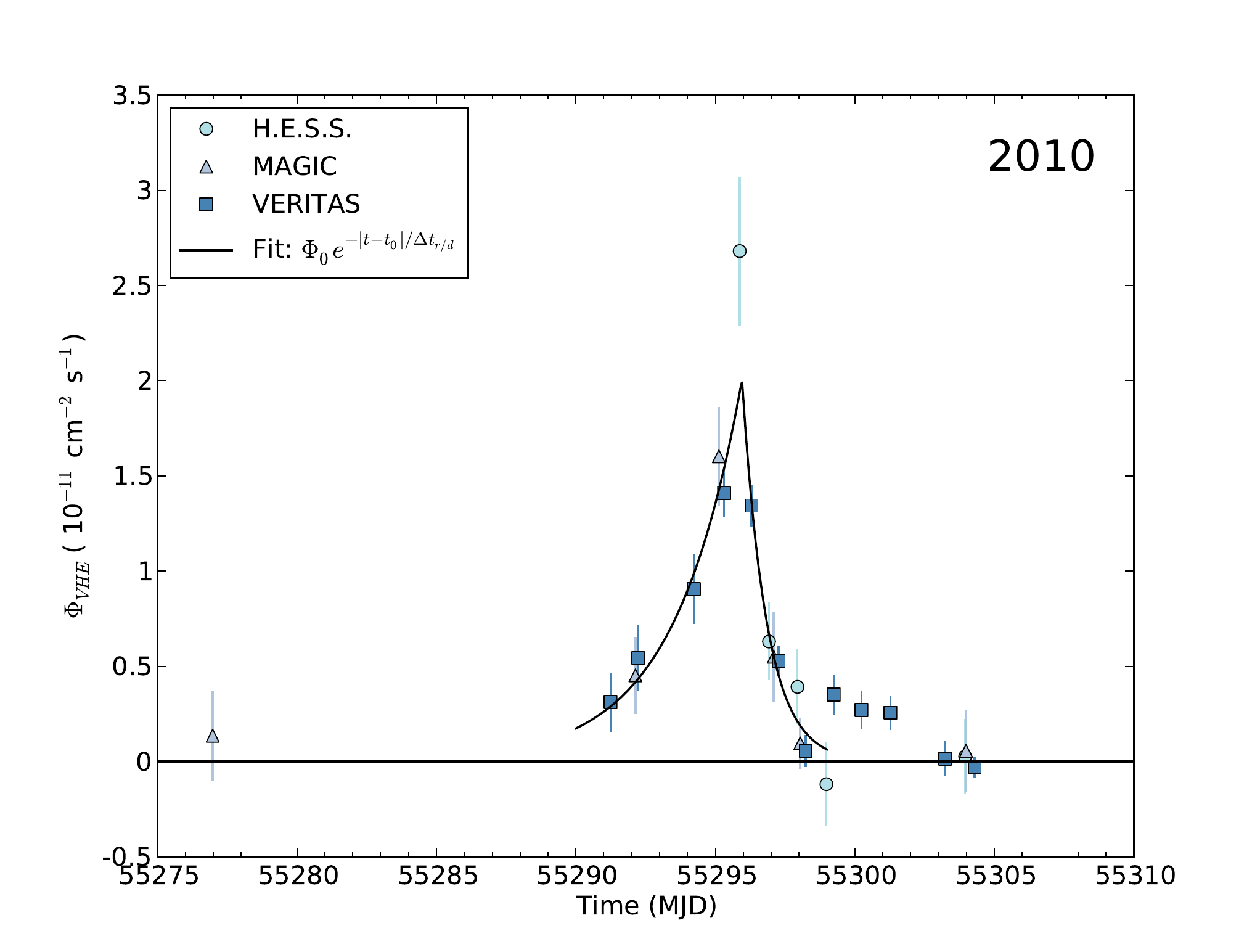}
  \caption{The 2010 April VHE flare of the radio galaxy M87 as observed 
  by different Cherenkov Telescopes. Significant day-scale activity is 
  evident. The curve shows a fit of an exponential function to the data. 
  From Ref.~\cite{Abramowski12}.}
\end{figure}
%%%%%%%%%%%%%%%%%%%%%%%%%%%%%%%%%%%

\noindent The Perseus Cluster galaxy {\it IC~310} has been detected by 
MAGIC above 300 GeV in about 21h of data (taken in 2009/2010) at an 
average level of $\sim 3\%$ of the Crab Nebula \cite{Aleksic10}. The 
source has been originally classified in the literature as a head-tail RG, 
but a recent radio analysis finds little evidence for jet bending \cite{Kadler12} 
(see also Ref. \cite{Rector99} for the proposal that IC~310 may instead be a 
weakly beamed blazar). The VHE spectrum measured between 150 GeV 
and 7 TeV is very hard (even harder than in M87) and compatible with a 
single power law of photon index $\Gamma \simeq 2.0$. There is clear 
evidence for VHE variability on yearly and monthly time scales, with some 
indications for day-scale activity found in a new analysis \cite{Eisenacher},
features that are reminiscent of the VHE activity seen in M87. Note however, 
that due to its larger distance, IC~310 needs to be intrinsically much more 
luminous.\\ 
In addition to IC~310, MAGIC has also recently reported the detection of 
the central dominant (FR I) cluster galaxy {\it NGC~1275} above $\sim
100$ GeV during enhanced high energy (Fermi-LAT) activity in 46h of 
data (taken between 08/2010-02/2011) \cite{Aleksic12}. This source is 
known to have radio jets misaligned by $\simgt30^{\circ}$. Fermi-LAT 
data indicate significant flaring activity above 0.8 GeV down to time 
scales of days \cite{Brown11}. At VHE energies the situation is less
evident, with no evidence for VHE variability found in the noted data set, 
while a recent, improved analysis of an earlier (10/2009-02/2010) VHE 
data set provides hints for a possible month-type variability \cite{Colin}. 
NGC1275 shows a very steep VHE spectrum (power law photon index 
$\Gamma \simeq 4.1$) extending up to $\sim500$ GeV. When compared 
with the hard HE (Fermi-LAT) spectrum (photon index $\Gamma\simeq 
2.1$), this implies the existence of a break or cut-off in the SED around 
some tens of GeV. 
%
 %%%%%%%%%%%   FIGURE 3  %%%%%%%%%%%%%%%%%
\begin{figure}[ht]
\includegraphics[height=.5\textheight]{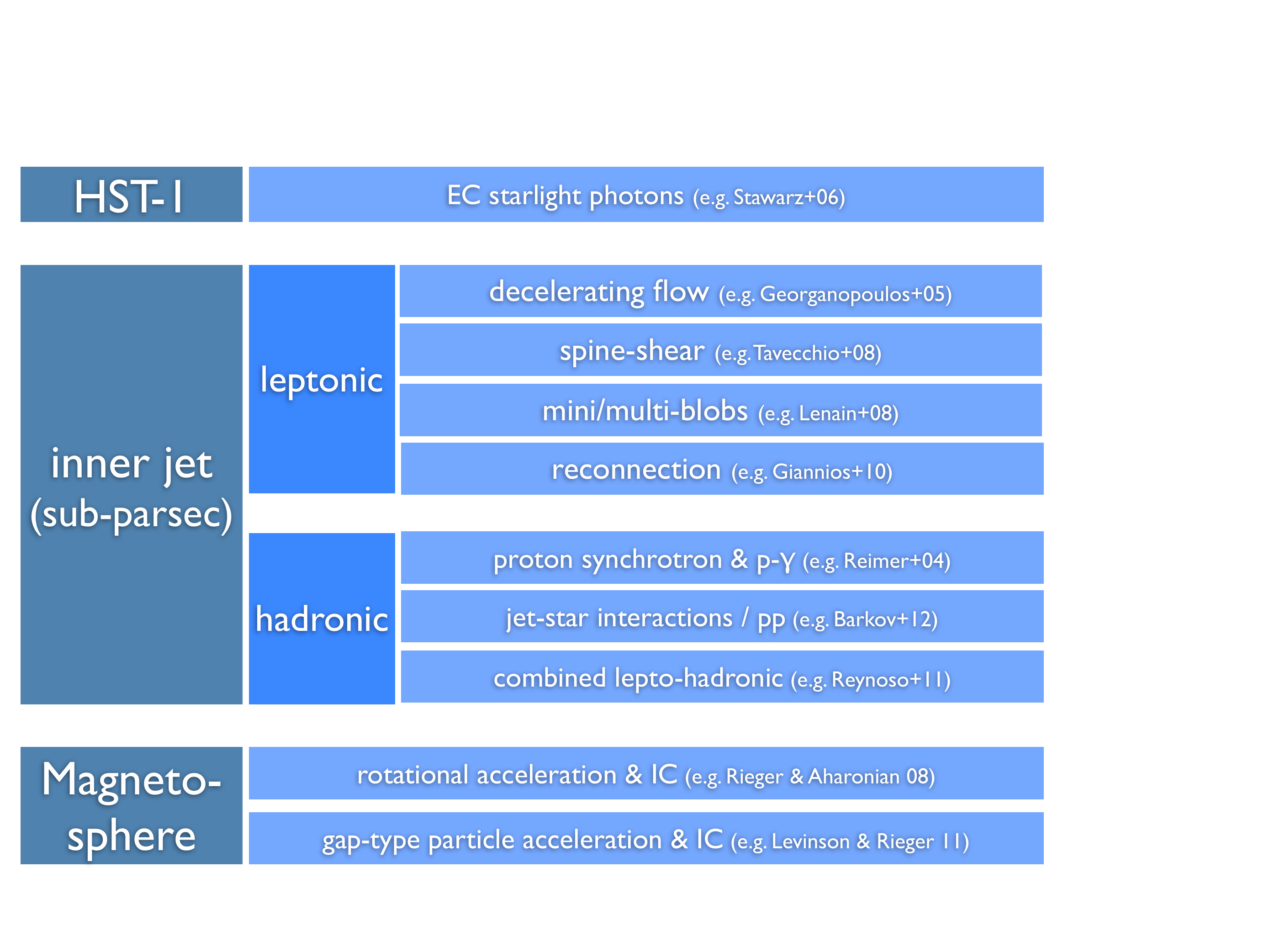}
\caption{Possible theoretical scenarios for the origin of the variable 
VHE emission as proposed for M87. For details, including some pros 
and cons of the models, see Ref. \cite{Rieger12b}.}
\end{figure}
%%%%%%%%%%%%%%%%%%%%%%%%%%%%%%%%%%

\section{On the origin of variable TeV gamma-rays from M87}
Being one of the best studied extragalactic object, M87 has become
a prominent benchmark for theoretical research. A multitude of VHE 
emission scenarios, operating on different scales and relying on 
either hadronic and/or leptonic radiation processes, have been 
developed and applied within recent years (cf. Fig. 3). The observed 
day-scale variability favors an origin of the variable TeV emission on 
scales of the inner jet and below, although outer-jet scenario may 
possibly contribute to the overall quiescent VHE source state (e.g., 
\cite{Rieger12b} for discussion).\\ 
(1) One recent theoretical development in the context of leptonic models
considers relativistic (Petschek-type) reconnection as acceleration 
mechanism of electrons to high energy \cite{Giannios10}, see Fig. 4 
for illustration. Compton up-scattering of external photons could then 
possibly account for the production of the observed, variable VHE 
gamma-rays \cite{Cui12}. Efficient reconnection in a highly magnetized 
($\sigma\sim 100$) electron-proton jet could lead to an additional 
relativistic velocity component of the ejected plasma ($\Gamma_r 
\simeq \sqrt{\sigma}$, at some angle to the main outflow direction) 
with respect to the mean bulk flow of the jet. This makes strong, 
differential Doppler boosting effects for the emitting zone possible, 
circumventing the problem of a modest Doppler factor for the general 
bulk flow. At present, there seem to be a number of open issues that 
need to be clarified within such an approach in order to assess its 
explanatory potential. This includes the fact that (i) typically much 
lower magnetizations ($\sigma\sim 10$) are expected for electron-proton 
(disk-driven) jets in AGN (in contrast to BZ-driven electron-positron
jets), that (ii) at the anticipated scale of VHE production a non-negligible 
magnetic guide field may persist allowing for weak dissipation only, 
and that (iii) the achievability of power-law electron acceleration much 
beyond the thermal Lorentz factor $\sqrt{\sigma} m_p/2m_e$ is not 
clear.\\
%
 %%%%%%%%%%%   FIGURE 4  %%%%%%%%%%%%%%%%%
\begin{figure}
\includegraphics[angle=270, width=7.5cm]{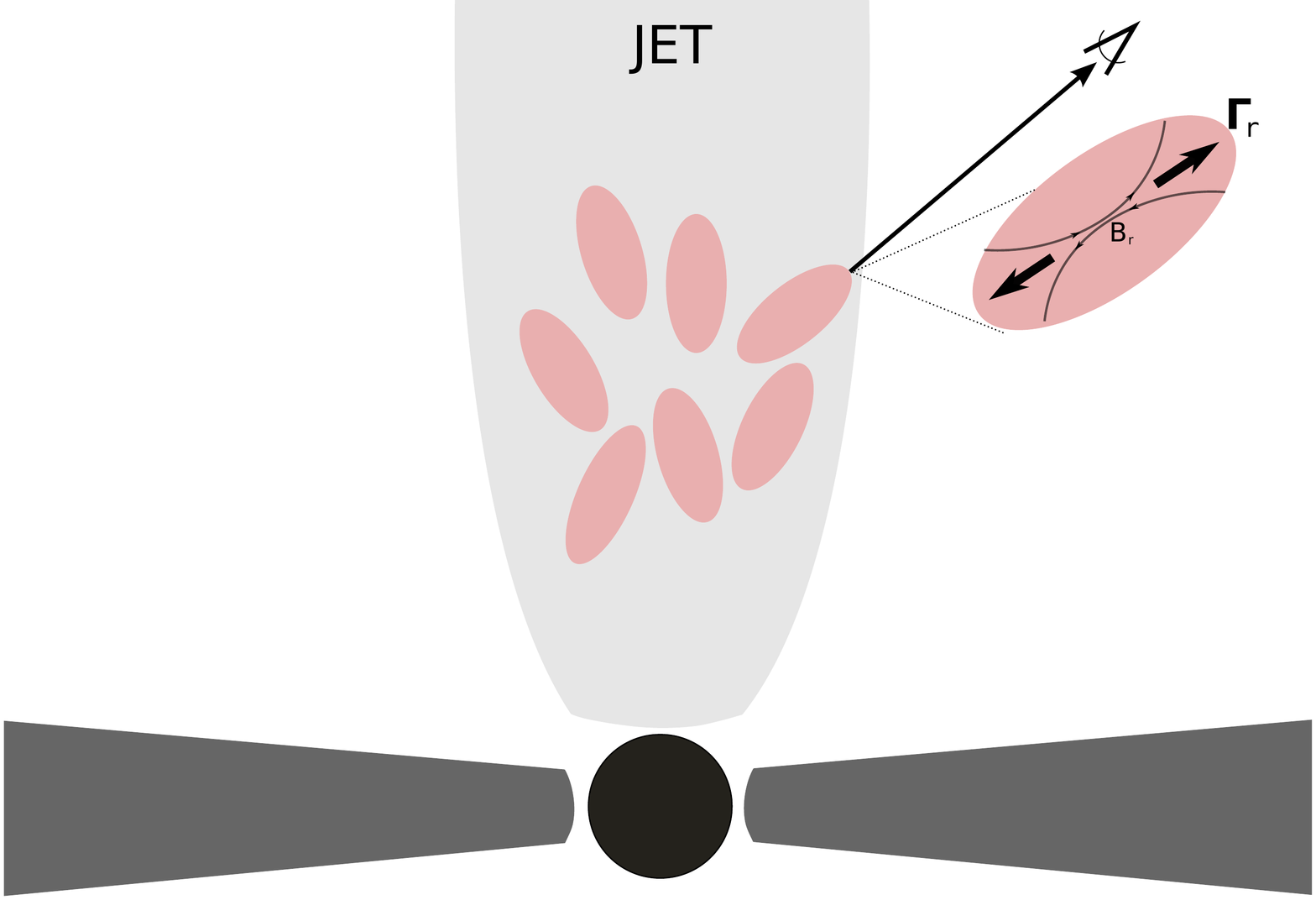}
\includegraphics[angle=270, width=7.5cm]{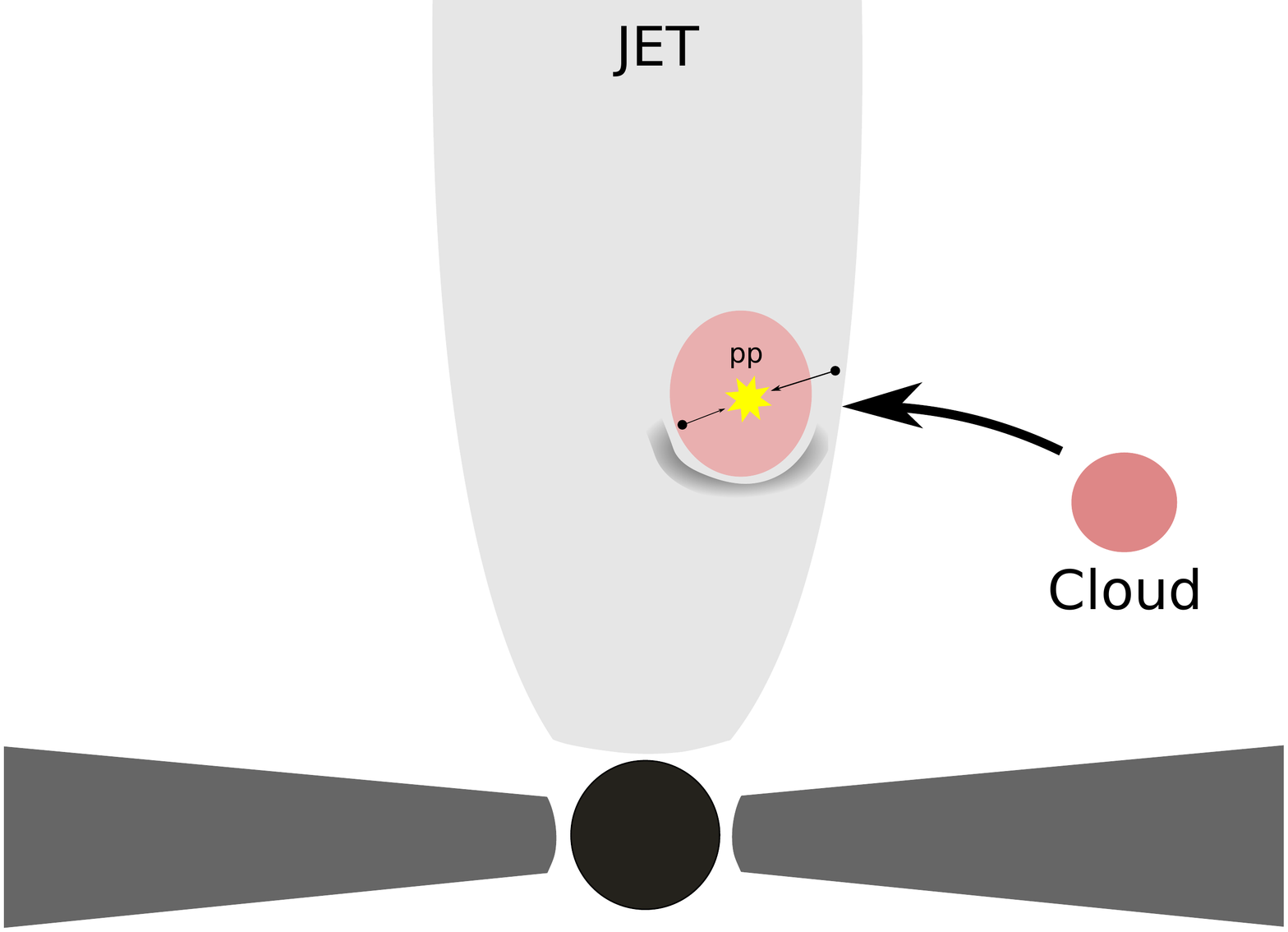}
\caption{LEFT: Illustration of a jets-in-jet scenario \cite{Giannios10}, 
where a variety of 'mini-jet' features are thought to be induced by 
relativistic magnetic reconnection events within the general jet flow. 
This could lead to an additional velocity component ($\Gamma_r$) 
relative to the main flow and allow a favorable orientation with
respect to the observer. RIGHT: Illustration of a hadronic model, 
where interactions of the jet with a massive gas cloud leads to 
Fermi (shock)-type proton acceleration and introduces a sufficient
target density to allow for efficient $pp$-collisions \cite{Barkov12}.}
\end{figure}
%%%%%%%%%%%%%%%%%%%%%%%%%%%%%%%%%%
\noindent(2) In the context of hadronic models, on the other hand, one new 
development considers inelastic proton-proton (pp) collisions (and $\pi^0$-decay) 
in the jet for the origin of the observed VHE emission. While AGN jets have 
been often considered not to carry (on average) a sufficient amount ($n_p$) 
of target matter \cite{Celotti98,Wardle98} to ensure the efficiency of this (pp) 
process (given its characteristic timescale $t_{pp}' \simeq 10^{15}/n_p'$ sec), 
it seems possible that interactions of a red giant star or a massive, dense gas 
cloud (size $r_c$) with the base of the jet (radius $r_j$) could occasionally 
introduce the required amount of matter and thereby potentially drive fast VHE 
activity \cite{Barkov10,Barkov12}. Recent calculations suggest that such a 
scenario could account for the observed temporal and spectral properties if 
the jet is powerful enough and a sufficiently large fraction ($\propto r_c^2/
r_j^2$) of it can be channeled into VHE $\gamma$-ray production. Given the 
observed large opening angle and transversal dimension of the radio jet in 
M87, models of this type require a spine-shear-type jet configuration with 
most of the energy flux concentrated into a narrow core.\\
(3) An alternative explanation has been pursued within magnetospheric models 
(see e.g. \cite{Rieger11} for review), motivated by the observed day-scale VHE 
variability and the observed radio-VHE connection in 2008. For a black hole 
mass in M87 expected to be in the range $M_{\rm BH} \simeq (2-6) \times 10^9\,
M_{\odot}$ \cite{Rieger12b}, anticipated variability  timescales could possibly be 
as small as $\sim r_g/c =0.1$ d, allowing to accommodate the observed rapid 
VHE variability. 
One of the latest developments in this context analyses the possible link between 
VHE $\gamma$-ray activity and jet formation \cite{Levinson11}. A schematic 
representation is shown in Fig. 5:\\  
% 
%%%%%%%%%%%   FIGURE   5 %%%%%%%%%%%%%%%%%%%%
\begin{figure}
\includegraphics[angle=270, width=.95\textwidth]{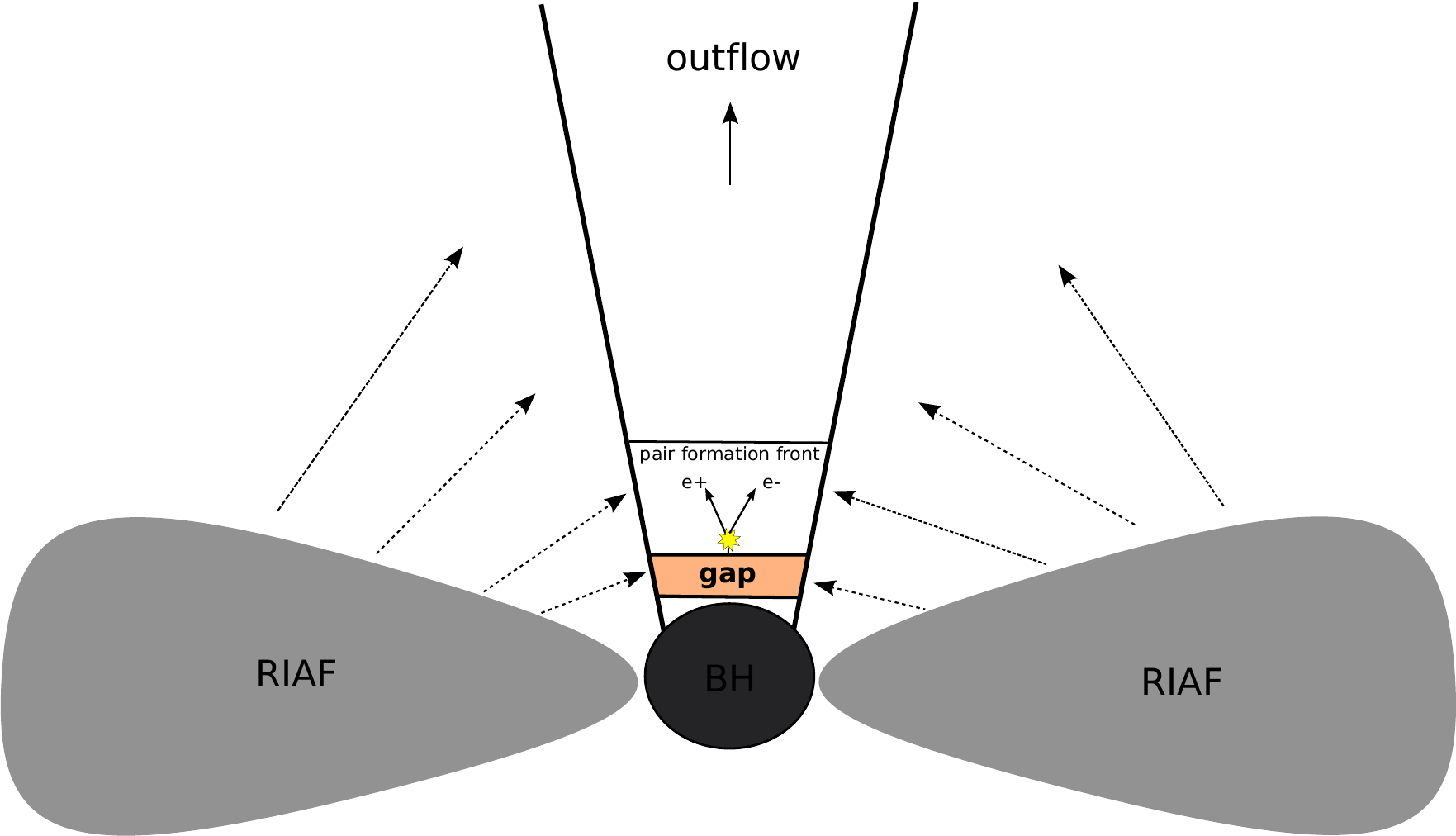}
\caption{VHE $\gamma$-ray production within a black hole-magnetospheric 
scenario \cite{Levinson11}: Electrons and positrons are accelerated in a 
vacuum gap to high Lorentz factors. The gap is exposed to soft radiation field 
of the disk. Curvature emission and inverse Compton up-scattering of ambient 
disk photons yields VHE $\gamma$-rays and triggers $e^+e^-$-cascades. 
The resultant large pair multiplicity establishes the force-free outflow seen 
in VLBA images.}
\end{figure}
%%%%%%%%%%%%%%%%%%%%%%%%%%%%%%%%%%%%%%
In this scenario, a maximally rotating black hole, which is embedded in an external 
magnetic field $B$ generates an electric potential difference across a (vacuum) gap 
of height $h$, given by $\Delta V_e=1.7\times10^{20} B_3 M_9(h/r_s)^2$ Volts, 
where $B_3=B/10^3$ G and $M_9=M_{\rm BH}/10^9~M_{\odot}$. Pair creation in 
a hot accretion flow (i.e. by $\gamma\gamma$-annihilation of MeV photons in an 
ADAF or RIAF) then leads to the injection of primary electrons and positrons into the 
gap, with the charged particles being quickly accelerated to high Lorentz factors. 
The number density of ADAF bremsstrahlung MeV photons is $n_{\gamma} \propto 
\dot{m}^2$, so that the $e^+e^-$-injection rate inside the magnetosphere roughly 
becomes 
$n_{\pm} \sim \sigma_{\gamma\gamma} n_{\gamma}^2 r_s \propto \dot{m}^4$. 
When expressed in terms of the Goldreich-Julian (GJ) density $n_{\rm GJ} =
\Omega B/(2\pi e c) \propto \dot{m}^{1/2}$, the ratio becomes \cite{Levinson11}
\[
n_\pm/n_{\rm GJ}\simeq 0.06~\dot{m}_{-4}^{7/2} M_9^{1/2}.
\]  This numerical estimate has some uncertainty as the bremsstrahlung spectrum 
is dependent on $T_e$, which itself depends on the accretion rate $\dot{m} \sim 
10^{-4}$ (expressed in terms of the Eddington rate). The sensitive dependence on 
$\dot{m}$ suggests, however, that a gap (with $n_\pm/n_{\rm GJ} < 1$) can be 
formed during periods of low accretion. Gap-type particle acceleration and 
subsequent curvature emission and, in particular, inverse Compton (IC) up-scattering 
of disk photons then produces $\gamma$-rays with a spectrum extending up to $10^4$ 
TeV. Different contributions dominating at different energy bands could then lead 
to a different variability behavior as e.g. seen by MAGIC during the 02/2008-VHE 
high state in M87 \cite{Albert10}. For low accretion rate, VHE photons with 
energies below several TeV can escape $\gamma\gamma$-absorption (see e.g., 
\cite{Levinson11,Rieger12b} for details). On the other hand, IC photons with energies 
above $\sim10$ TeV interact with the ambient radiation field initiating pair cascades 
just above the gap. This leads to a large pair multiplicity ($M=n_{2\pm}/n_{\rm GJ}$
up to $10^3$). A force-free outflow becomes established above the pair formation 
front and accounts for the jet feature seen in radio VLBA. Given the sensitive 
dependence of gap formation (primary pair injection) and photon escape (optical 
depth for $\gamma\gamma$-absorption) on accretion rate, any emission from the 
gap is likely to be intermittent. Modest changes in accretion rate, for example, may 
naturally give rise to the variability of the TeV emission and to fluctuations of the 
resultant force-free outflow. 

\section{Conclusions}
Radio galaxies have emerged as a new and particularly interesting VHE source 
class on the extragalactic sky. Evidence for $\gamma$-ray variability, as seen in 
most of the sources (apart from Cen~A) suggests that the observed $\gamma-$ray 
emission originates in a compact region. In reality, the situation may be somewhat 
more complex with extended regions contributing to the average/low states of the 
source, and different (additional) zones dominating during high VHE source states. 
In the case of M87, the observed VHE flaring characteristics (such as, e.g., the hard 
$\gamma$-ray spectrum extending beyond 10 TeV and the day-scale VHE activity) 
are of fundamental interest as they allow us to probe the near black hole environment 
and to study the possible link between VHE activity and jet formation in AGN. Radio 
galaxies like M87 thus appear to be particularly promising targets for future \& more 
sensitive instruments like CTA.

 \begin{theacknowledgments}
{Discussions with Felix Aharonian, Amir Levinson, Maxim Barkov, Matthias Beilicke 
and Martin Raue, and financial support by a LEA Fellowship are gratefully 
acknowledged.} 
\end{theacknowledgments}

\bibliographystyle{aipproc}   % if natbib is available

\end{document}